\documentstyle[preprint,aps,epsfig]{revtex}
\tightenlines
\begin{document}
%
%
%
\draft
%
%
\title{
$I=2$ Pion Scattering Phase Shift with Wilson Fermions
}
%
%
\author{
       S.~Aoki,$^{\rm 1    }$
   M.~Fukugita,$^{\rm 2    }$
  S.~Hashimoto,$^{\rm 3    }$
 K-I.~Ishikawa,$^{\rm 1, 4 }$
   N.~Ishizuka,$^{\rm 1, 4 }$
    Y.~Iwasaki,$^{\rm 1    }$
     K.~Kanaya,$^{\rm 1    }$
     T.~Kaneko,$^{\rm 3    }$
  Y.~Kuramashi,$^{\rm 3    }$
       V.~Lesk,$^{\rm 4    }$
      M.~Okawa,$^{\rm 5    }$
  Y.~Taniguchi,$^{\rm 1    }$
      A.~Ukawa,$^{\rm 1, 4 }$
 T.~Yoshi\'{e},$^{\rm 1, 4 }$ \\
(CP-PACS Collaboration)
}
\address{
${}^{\rm 1}$ Institute of Physics, University of Tsukuba            , Tsukuba, Ibaraki 305-8571, Japan \\
${}^{\rm 2}$ Institute for Cosmic Ray Research, University of Tokyo , Tanashi, Tokyo   188-8502, Japan \\
${}^{\rm 3}$ High Energy Accelerator Research Organization (KEK)    , Tsukuba, Ibaraki 305-0801, Japan \\
${}^{\rm 4}$ Center for Computational Physics, University of Tsukuba, Tsukuba, Ibaraki 305-8577, Japan \\
${}^{\rm 5}$ Department of Physics, Hiroshima University, Higashi-Hiroshima, Hiroshima 739-8526, Japan \\
}
%
\date{ \today }
\maketitle
%
%
\setlength{\baselineskip}{18pt}
%
%
\begin{abstract}
We present a lattice QCD calculation of the scattering phase shift for the
$I=2$ $S$-wave two-pion system using the finite size method proposed by L\"uscher.
We work in the quenched approximation employing the standard plaquette
action at $\beta=5.9$ for gluons and the Wilson fermion action for quarks.
The phase shift is extracted from the energy eigenvalues of the two-pion system,
which are obtained by a diagonalization of the pion
4-point function evaluated for a set of relative spatial momenta.
In order to change momentum of the two-pion system, calculations are carried out on
$24^3\times 60$,
$32^3\times 60$, and
$48^3\times 60$ lattices.
The phase shift is successfully calculated
over the momentum range $0 < p^2 < 0.3\ {\rm GeV}^2$.
\end{abstract}
%
%
%
%
\narrowtext
%
%
\section{ Introduction }
Calculation of scattering phase shift
is an important step for expanding our understanding
of strong interactions based on lattice QCD beyond the hadron mass spectrum.
For scattering lengths, which are the threshold values of phase shifts,
several studies have already been carried out.
For the simplest case of the two-pion system,
the $I=2$ scattering length has been calculated in detail
\cite{SGK,Kuramashi,GPS,Alford,JLQCD,JLQCD.new,LZCM}
including the continuum extrapolation \cite{JLQCD,JLQCD.new,LZCM}.
There is also a pioneering attempt at the $I=0$ scattering length~\cite{Kuramashi},
which is much more difficult due to the presence of box and disconnected contributions.
For the scattering phase shift, in contrast,
there has been only one calculation for $I=2$ by Fiebig {\it et.al.},
who used lattice simulations to estimate the effective two-pion potential
and used it to calculate the phase shift in a quantum mechanical treatment~\cite{Fiebig}.

In this article,
we calculate the $I=2$ $S$-wave two-pion scattering phase shift
applying the L\"uscher's finite size method~\cite{Luscher_1,Luscher_2}.
Technically the key feature is the extraction
of the two-pion energy eigenvalues from the pion $4$-point function.
This is successfully solved by a diagonalization method
proposed by L\"uscher and Wolff~\cite{Luscher-Wolff}
for $O(3)$ non-linear $\sigma$ model in $2$-dimensions.
We also extract the scattering length from the phase shift data,
and compare it with previous calculations.
We work in quenched lattice QCD employing the standard plaquette
action for gluons and the Wilson fermion action for quarks.

We wish to mention that the study of the two-pion scattering phase shift
also has important impact on the calculation of the $K\to\pi\pi$ decay amplitudes.
A direct calculation of the amplitude from the $4$-point function
$\langle 0 | \pi(t_\pi) \pi(t_\pi) H_W(t_H) K(t_K) | 0 \rangle$
is very difficult, as pointed out by Maiani and Testa~\cite{MT-nogo},
because the 4-point function at large times is dominated
by the two-pion ground state with zero relative momenta,
which differs from the final state of the decay having a non-zero relative momentum.
An exception is the amplitude from the $K$ meson to the two-pion ground state itself,
because this can be calculated by taking the two-pion state
with zero relative momentum in the final state.
However, the amplitude thus obtained is unphysical,
and a reconstruction of the physical amplitude
using some effective theory of QCD,
for example chiral perturbation theory (CHPT), is needed.
Using such an effective theory causes large uncertainties
in the lattice prediction of the decay amplitude.
Hence, a method for direct calculation of the $K\to\pi\pi$ decay amplitude
has been strongly desired.

Recently Lellouch and L\"uscher~\cite{Lellouch-Luscher}
obtained a relation between the lattice and the physical amplitude
in the two-pion center of mass system with the energy $E_\pi = m_K$.
In their derivation no effective theory is used.
Lin {\it et.al.}~\cite{LMST} derived the relation from a different approach,
and extended it to the general two-pion system with the energy $E_\pi \not= m_K$.
They also investigated the limitation of the relation.

In order to apply the relation to obtain the physical decay amplitude,
one has to calculate the amplitude from $K$ meson to the two-pion energy
eigenstate with non-zero momenta on the lattice.
This is the same problem as one encounters in the calculation of phase shifts
using the L\"uscher's method.
Thus study of the two-pion system represents a first step toward $K\to\pi\pi$ decay.

This paper is organized as follows.
In Sec.~\ref{TWO_S}
we describe the formalism for calculation of the scattering length
and phase shift~\cite{Luscher_1,Luscher_2}.
We also discuss the method of extraction of energy eigenvalues
of the two-pion system from the pion $4$-point functions.
The simulation parameters used in this work are given in Sec.~\ref{THREE_S}.
In Sec.~\ref{FOUR_S}
we analyze the behavior of the $4$-point functions,
and show that the diagonalization technique proposed by L\"uscher and Wolff allows
to extract the energy eigenvalues.
We then present results for the pion phase shift.
Our conclusions are given in Sec.~\ref{FIVE_S}.
A preliminary report of the present work was presented
in Ref.~\cite{CP-PACS.PHSH.old}.
%
%
\section{\label{TWO_S} Methods }
\subsection{ Finite Size Method }
The energy eigenvalues
of a non-interacting two-pion system
on a finite periodic box of a size $L^3$ are quantized as follows :
\begin{equation}
  E_n
    = 2 \cdot \sqrt{ m_\pi^2 + p_n^2 }           \ ,
        \quad  p_n^2 = ( 2 \pi / L)^2 \cdot n    \ ,
        \quad  n \in {\bf Z}
\ .
\label{EIGEN_E_NOINT_eq}
\end{equation}
In the interacting case the $n$-th energy eigenvalue is given by
\begin{equation}
  \bar{E}_n
    = 2 \cdot \sqrt{ m_\pi^2 + \bar{p}_n^2 }              \ ,
        \quad \bar{p}_n^2 = ( 2 \pi / L)^2 \cdot \bar{n}  \ ,
        \quad \bar{n} \not\in {\bf Z}                     \ ,
        \quad n \in {\bf Z}
\ .
\label{EIGEN_E_INT_eq}
\end{equation}
The energy eigenvalue is written as that of the non-interacting two-pion
system with momentum $\bar{\bf p}_n$ and $-\bar{\bf p}_n$,
but the quantity $\bar{n} = L^2 / (2\pi)^2 \cdot \bar{p}_n^2$ is not an integer.
The momentum $\bar{p}^2_n$ satisfies the L\"uscher relation~\cite{Luscher_1,Luscher_2},
\begin{equation}
  \tan \delta( \bar{p}_n )
    = \frac{ \pi^{3/2} \sqrt{ \bar{n} } }{ {\cal Z}_{00} ( 1 ; \bar{n} ) }
\ ,
\label{LUSCHER_eq}
\end{equation}
where $\delta(\bar{p}_n)$ is the $S$-wave scattering phase shift at infinite volume
and
\begin{equation}
     {\cal Z}_{00}( k ; \bar{n} )
   = \frac{ 1 }{ \sqrt{ 4\pi } }
        \cdot  \sum_{ {\bf m} \in {\bf Z}^3 } ( m^2 - \bar{n} )^{-k}
\ .
\label{ZETA.I_eq}
\end{equation}
Using (\ref{LUSCHER_eq}),
we can obtain the scattering phase shift from the energy eigenvalue
calculated in lattice simulations.
The scattering length is given by
$a_0 = \lim_{\bar{p}\to 0} \tan \delta(\bar{p}) / \bar{p}$.

In the limit of large volume or weak two-pion interactions,
we find
\begin{equation}
 \bar{p}_n^2 - p_n^2 = O(1/L^3)
\qquad \mbox{or} \qquad
 \bar{n} - n = O(1/L)
\end{equation}
from (\ref{LUSCHER_eq}) and (\ref{ZETA.I_eq}).
Therefore, taking the volume $L^3$ to be large in lattice calculations,
we can employ an expansion of ${\cal Z}_{00} ( 1 ; \bar{n} )$
around $n \in {\bf Z}$ given by
\begin{equation}
    \sqrt{ 4\pi } \cdot {\cal Z}_{00}( 1 ; \bar{n} )
    =  - \frac{ N_n }{ \bar{n} - n }
       + \lim_{N\to\infty} \sum_{j=1}^{N} Z_{00} ( j ; n ) \cdot ( \bar{n} - n )^{j-1}
\label{ZETA.II_eq}
\ ,
\end{equation}
where
\begin{equation}
    Z_{00} ( j ; n )
      = \lim_{ \bar{n} \to n }
             \biggl[
                   \sqrt{4\pi} \cdot {\cal Z}_{00}( j ; \bar{n} )
                -  N_n \cdot ( n - \bar{n} )^{-j}
             \biggr]
\label{ZETA.III_eq}
\end{equation}
and $N_n = \sum_{ {\bf m} \in {\bf Z}^3 } \delta( m^2 - n )$.
In this work we use the expansion (\ref{ZETA.II_eq}) with $N=10$,
with which the numerical error for all our simulation parameters
are under $O(10^{-8})$.
The numerical calculation of $Z_{00}(j;n)$ is discussed in Ref.~\cite{Luscher_1}.
The values for several $j$'s and $n$'s are tabulated in Table~\ref{ZETA.table}.
%
%
\subsection{ Extraction of Energy Eigenvalues of Two-Pion System }
In order to obtain the energy eigenvalues of the two-pion system
we construct the pion $4$-point function :
\begin{equation}
  G_{nm}^{(N_R)}(t) = \langle 0 | \Omega_n ( t ) \Omega_m^{(N_R)} ( t_S ) | 0 \rangle
\ .
\label{GPK_I}
\end{equation}
Here $\Omega_n( t )$ is an interpolating field for the $S$-wave two-pion
system at time $t$ given by
\begin{equation}
  \Omega_n( t )
        = \frac{ 1 }{48} \cdot
                \sum_{R}
                      \pi(  R(\vec{p}_n), t )
                      \pi( -R(\vec{p}_n), t )
\ ,
\end{equation}
where $\pi( \vec{p}_n, t)$ is the pion interpolating field
with lattice momentum $\vec{p}_n$ at time $t$.
The vector $\vec{p}_n$ satisfies $p_n^2 = (2\pi/L)^2 \cdot n$ ($n \in {\bf Z}$),
and $R$ is an element of cubic group which has  48 elements.
The summation over $R$ is the projection to the ${\bf A^{+}}$ sector of
the cubic group,
which equals the $S$-wave state in the continuum
ignoring effects from states with angular momentum $L \geq 4$.

For source we use another operator $\Omega_n^{(N_R)} ( t )$
defined by
\begin{equation}
\Omega_{n}^{(N_R)}( t )
   = \frac{1}{N_R} \cdot
     \sum_{j=1}^{N_R}
       \pi(   \vec{p}_n, t,  \xi_j )
       \pi( - \vec{p}_n, t, \eta_j )
\ ,
\end{equation}
where
\begin{equation}
\pi( \vec{p}_n, t, \xi_j  )
      = \frac{1}{L^3} \cdot
          \biggl[
             \sum_{\vec{x}}   \bar{ q } ( \vec{x} , t )
                {\rm e}^{  i \vec{p}_n \cdot \vec{x} }
                        \cdot \xi_j^\dagger (\vec{x})
        \biggr]
    \gamma_5
        \biggl[
           \sum_{\vec{y}}          q ( \vec{y} , t )
                       \cdot \xi_j (\vec{y})
         \biggr]
\label{OPPI_S.eq}
\ .
\end{equation}
The field $\pi( \vec{p}_n, t, \eta_j )$ is defined as $\pi( \vec{p}_n, t, \xi_j )$
by changing $\xi_j(\vec{x})$ to $\eta_j(\vec{x})$.
The functions $\xi_j(\vec{x})$ and $\eta_j(\vec{x})$ are
orthogonal complex random numbers in $3$-dimensional space, whose property is
\begin{equation}
   \lim_{ N_R \to \infty }
   \frac{1}{N_R} \cdot
   \sum_{j=1}^{N_R}
       \xi_j^\dagger (\vec{x})
       \xi_j         (\vec{y})
                = \delta^3 ( \vec{ x } - \vec{ y } )
\ .
\end{equation}
The pion $2$-point function is constructed as
\begin{equation}
  G_n^{\pi (N_R)} (t)
   = \frac{1}{N_R} \cdot \sum_{j=1}^{N_R}
     \ \langle 0 | \pi( \vec{p}_n, t ) \pi( - \vec{p}_n , t_S , \xi_j ) | 0 \rangle
\ .
\end{equation}
When the number of random noise source $N_R$ is taken large
or the number of gauge configurations becomes large, we expect
\begin{eqnarray}
          G_{nm}^{(N_R)}(t)
& \sim &  G_{nm}(t)
    =      \langle 0 | \Omega_n ( t ) \Omega_m ( t_S ) | 0 \rangle
\ ,  \cr
          G_n^{\pi(N_R)} (t)
& \sim &  G_n^{\pi}(t)
    =     \langle 0 | \pi( \vec{p}_n, t ) \pi( -\vec{p}_n, t_S ) | 0 \rangle
\ ,
\label{GPK_II}
\end{eqnarray}
and the $4$-point function will be symmetric under exchange of the
sink and source momenta.
In our numerical calculations
we use ${\rm U}(1)$ random numbers and take $N_R=2$.
The number of configurations is $200$, $286$ and $52$ depending on
the lattice size as shown in Sec.~\ref{THREE_S}.
We always check the symmetry of the $4$-point function across the midpoint
in the temporal direction before analysis.

The $4$-point function can be rewritten in terms
of the energy eigenvalue $\bar{E}_j$ and eigenstate 
$| \bar{\Omega}_j \rangle$ as
\begin{equation}
  G_{nm}(t) = \sum_{ j \in {\bf Z} }
         \ \frac{   \langle 0  |  \Omega_n  |   \bar{\Omega}_j   \rangle
                    \langle \bar{\Omega}_j  | \Omega_m | 0 \rangle
                }{ \langle \bar{\Omega}_j | \bar{\Omega}_j \rangle }
    \cdot  {\rm e}^{ - \bar{E}_j \cdot ( t - t_S ) }
\ ,
\label{GPK_III}
\end{equation}
where $\Omega_n = \Omega_n(0)$ and we assume non-degeneracy of energy eigenstates.
The $j$-th energy $\bar{E}_j = 2 \cdot \sqrt{ m_\pi^2 + \bar{p}_j^2 }$
satisfies the L\"uscher relation (\ref{LUSCHER_eq}).
Since the matrix element $\langle 0 | \Omega_n | \bar{\Omega}_m \rangle$
is not diagonal generally,
the $4$-point function $G_{nm}(t)$ contains many exponential terms
and is not a diagonal matrix with respect to the momentum indices $n$ and $m$.
For simplicity we introduce the following matrices :
\begin{eqnarray}
&&  V_{nm} =
        \langle 0 | \Omega_n | \bar{\Omega}_m \rangle
        / \sqrt{  \langle \bar{\Omega}_m | \bar{\Omega}_m \rangle }
     \cr
&&  \Delta_{nm}(t)
     = \delta_{nm}
       \cdot {\rm e}^{ - \bar{E}_n \cdot ( t - t_S ) }
\label{defof_VD}
\end{eqnarray}
and rewrite the $4$-point function in the following matrix form.
\begin{equation}
  G(t) = V \ \Delta(t) \ V^{T}
\ ,
\label{GPK_IV}
\end{equation}
where $n$ and $m$ are regarded as matrix indices.

The extraction of the energy eigenvalues from multi-exponential Green function
such as (\ref{GPK_IV}) is non-trivial.
One can attempt multi-exponential fitting to extract them,
but it is very difficult in general.
A method of extraction was proposed by L\"uscher and Wolff~\cite{Luscher-Wolff}.
They applied it to the $O(3)$ non-linear $\sigma$ model in $2$-dimensions
and obtained the scattering phase shift.
This method has been used for many statistical systems~\cite{LW_stat}
and also for the $I=2$ two-pion system of QCD~\cite{Fiebig}.
In their method the following matrix is diagonalized at each $t$, 
\begin{equation}
   M(t,t_0) = G(t_0)^{-1/2} \ G(t) \ G(t_0)^{-1/2}
\ ,
\label{M_I}
\end{equation}
where $t_0$ is some reference time.
The eigenvalues $\lambda(t,t_0)$ of $M(t,t_0)$ can be obtained
easily from (\ref{GPK_IV}) and (\ref{M_I}) by
\begin{eqnarray}
&&   \lambda(t,t_0)
        = {\rm Ev} \biggl[ M(t,t_0) \biggr]
        = {\rm Ev} \biggl[ G(t) G(t_0)^{-1} \biggr]  \cr
&& = {\rm Ev} \biggl[ V \Delta(t) \Delta(t_0)^{-1} V^{-1} \biggr]
   = {\rm Ev} \biggl[ \Delta(t) \Delta(t_0)^{-1} \biggr]  \cr
&&   = \biggl\{ \exp \left( - \bar{E}_j \cdot ( t - t_0 ) \right)
                \ \biggl| \ j = 0, 1, 2 \cdots \biggr\}
\ .
\end{eqnarray}
Therefore after diagonalization of $M(t,t_0)$
we can obtain the energy eigenvalues $\bar{E}_j$ by a single exponential fitting.

%

In actual calculations
we can not calculate all the components of the $4$-point function precisely.
We have to set a momentum cut-off $p^2_{\rm cut} = (2\pi/L)^2\cdot N$.
Here we expect that the components of $G_{nm}(t)$ for $n,m \leq k $ are dominant
for the $k$-th eigenvalue $\lambda_k(t)$ in the large $t$ and $t_0$ region,
while the components $n,m > k$ are less important.
In this work we set $t_0$ and $t$ large
and investigate the cut-off dependence for $N \geq k$ .
%
%
\section{\label{THREE_S} Simulation Parameters }
Our simulation is carried out in quenched lattice QCD employing
the standard plaquette action for gluons at $\beta=5.9$
and the Wilson action for quarks.
Quark masses are chosen to be the same as in the previous study
of the quenched hadron spectroscopy by CP-PACS~\cite{CP-PACS.LHM},
{\it i.e.,}
$\kappa       = 0.1589  $, $0.1583  $, $0.1574  $, and $0.1566  $, which correspond to
$m_\pi/m_\rho = 0.491(2)$, $0.593(1)$, $0.692(1)$, and $0.752(1)$.
The lattice cut-off is estimated from the $\rho$ meson mass,
and equals $1/a=1.934(16) {\rm GeV}$.

\begin{figure}
\centerline{ \epsfig{ file=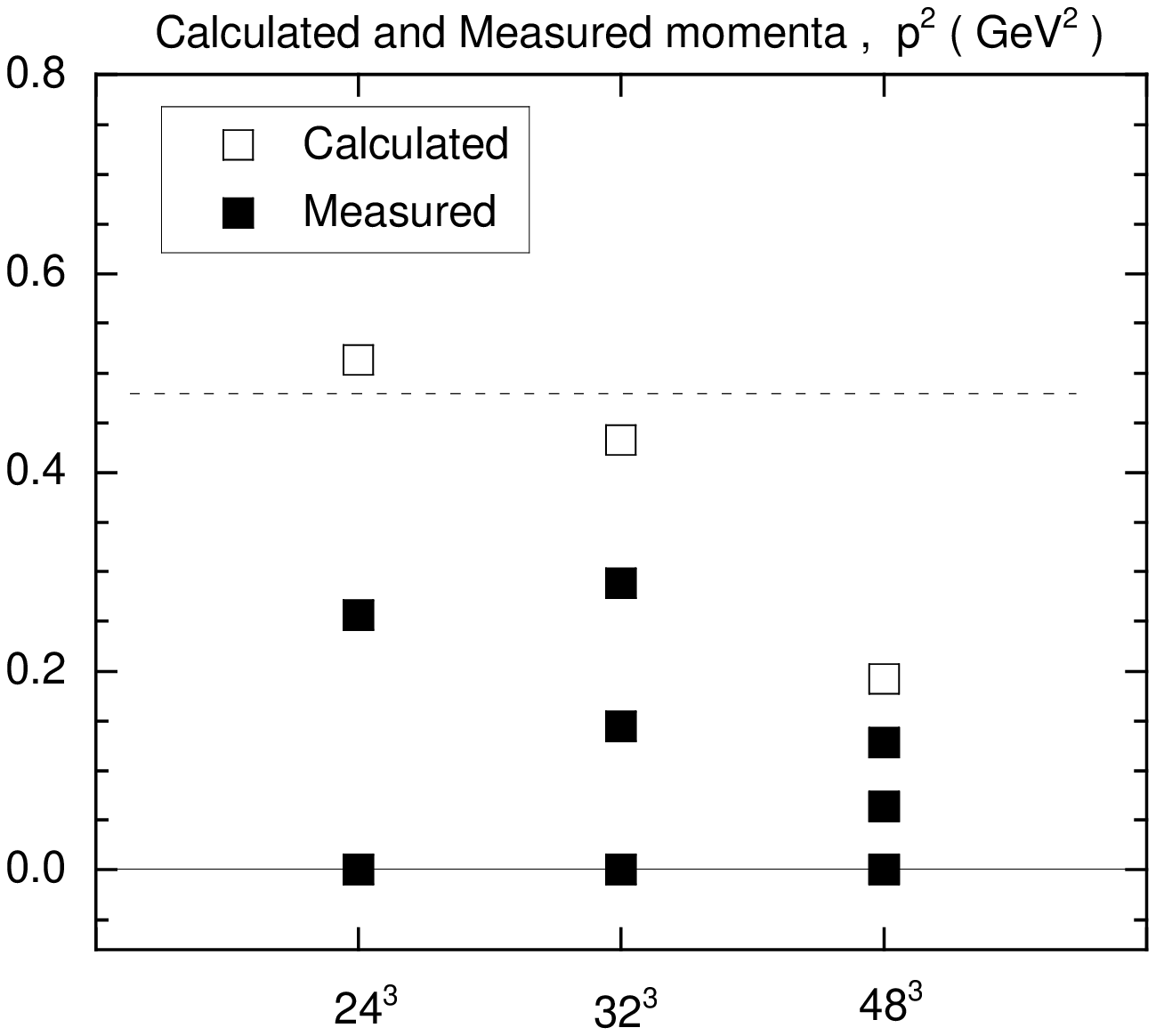, width=11.0cm } }
\vspace{0.5cm}
\caption{ \label{mom.fig}
Momenta in units of ${\rm GeV^2}$ used in this work for each lattice size.
We obtain scattering length and phase shift at the filled symbols.
Momenta marked by open symbols are used only to examine the momentum cut-off effects.
The broken line shows the upper limit of elastic scattering
for the smallest $m_\pi$ in this work, {\it i.e.}
$E = 2\cdot \sqrt{m_\pi^2 + p^2 } < 4 m_\pi$.
}
\end{figure}

In order to examine finite-size effects for the scattering length
and to change the momentum for the phase shift,
lattice simulations are carried out for three lattice sizes
with a fixed temporal size $T=60$.
The number of configurations and the momentum $p_n^2=(2\pi/L)^2\cdot n$
for each lattice size are tabulated below.
\begin{eqnarray}
&&   L^3  \qquad  \mbox{configurations}  \qquad  \mbox{  $n$        } \cr
&&  24^3  \qquad      \mbox{ 200 } \qquad \qquad \mbox{ \underbar{ 0 } , \underbar{ 1 } , \ 2      } \cr
&&  32^3  \qquad      \mbox{ 286 } \qquad \qquad \mbox{ \underbar{ 0 } , \underbar{ 1 } , \underbar{ 2 } , \ 3 } \cr
&&  48^3  \qquad \ \, \mbox{  52 } \qquad \qquad \mbox{ \underbar{ 0 } , \underbar{ 1 } , \underbar{ 2 } , \ 3 }
\ .
\end{eqnarray}
Here we calculate the phase shift
at the momenta marked by under-bar; those un-marked
are used to examine the momentum cut-off effects.
The momenta in units of ${\rm GeV}^2$ chosen in this work
are plotted in Fig.~\ref{mom.fig}.

We note that the two-pion energy eigenstates are
not degenerate for $n \leq 6$.
Since the effects from the states $n > 6$ can be thought to be negligible
for the first several low-energy states,
the non-degeneracy assumption in the derivation of the diagonalization method
in the previous section is justified.

Gluon configurations are generated with the $5$-hit heat-bath algorithm
and the over-relaxation algorithm mixed in the ratio of $1:4$.
The combination is called a sweep and
we skip $200$ sweeps between measurements of physical quantities.
Quark propagators are solved with the Dirichlet boundary condition imposed
in the time direction and the source operator set
at $t_S=8$ to avoid effects from the temporal boundary.
%
%
\section{ \label{FOUR_S} Results }
%
%
\subsection{ Effects of Diagonalization }
\begin{figure}
\centerline{ \epsfig{ file=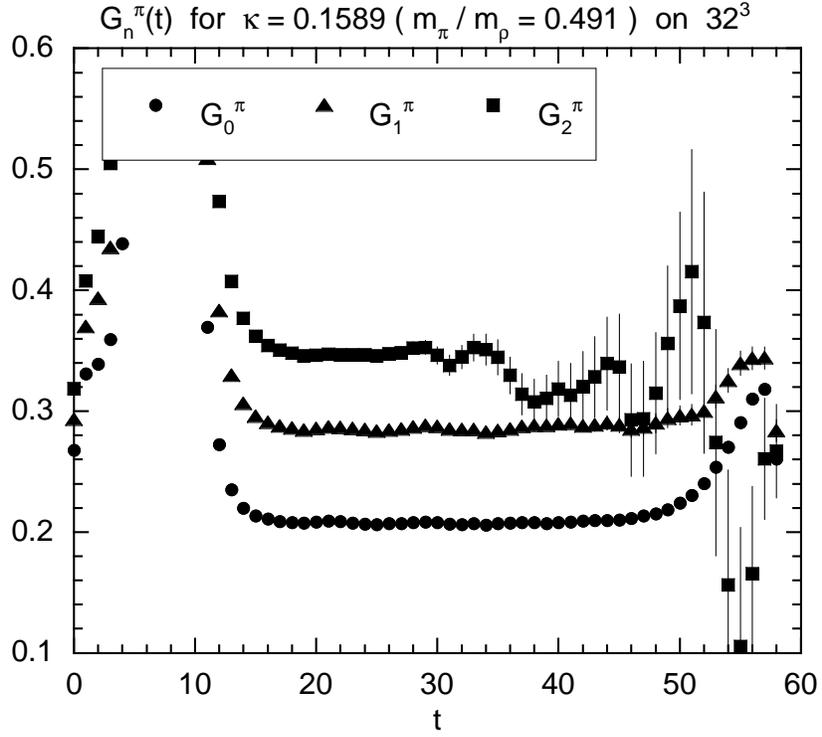, width=11.0cm } }
\vspace{0.5cm}
\caption{ \label{GP_PI1.fig}
Examples of effective mass of pion propagator $G_n^{\pi}(t)$
at $\kappa=0.1589$ ($m_\pi/m_\rho=0.491$) on a $32^3$ lattice.
The subscript $n$ refers to the momentum $p^2 = (2\pi/L)^2\cdot n$.
The source is located at $t=8$.
}
\end{figure}

In Fig.~\ref{GP_PI1.fig}
we show examples of effective mass of the pion propagator $G_n^{\pi}(t)$
for momenta $n=0, 1, 2$ ( $p^2=(2\pi/L)^2\cdot n$ )
at $m_\pi / m_\rho = 0.491$ on a $32^3$ lattice.
The source operator is located at $t_S=8$.
We observe a clear plateau over the time range $t \sim 18 - 46$ for small momenta,
but the signal becomes noisier for large momenta.
We also find very large effects from the temporally boundary for $t > 46$.

\begin{figure}
\centerline{ \epsfig{ file=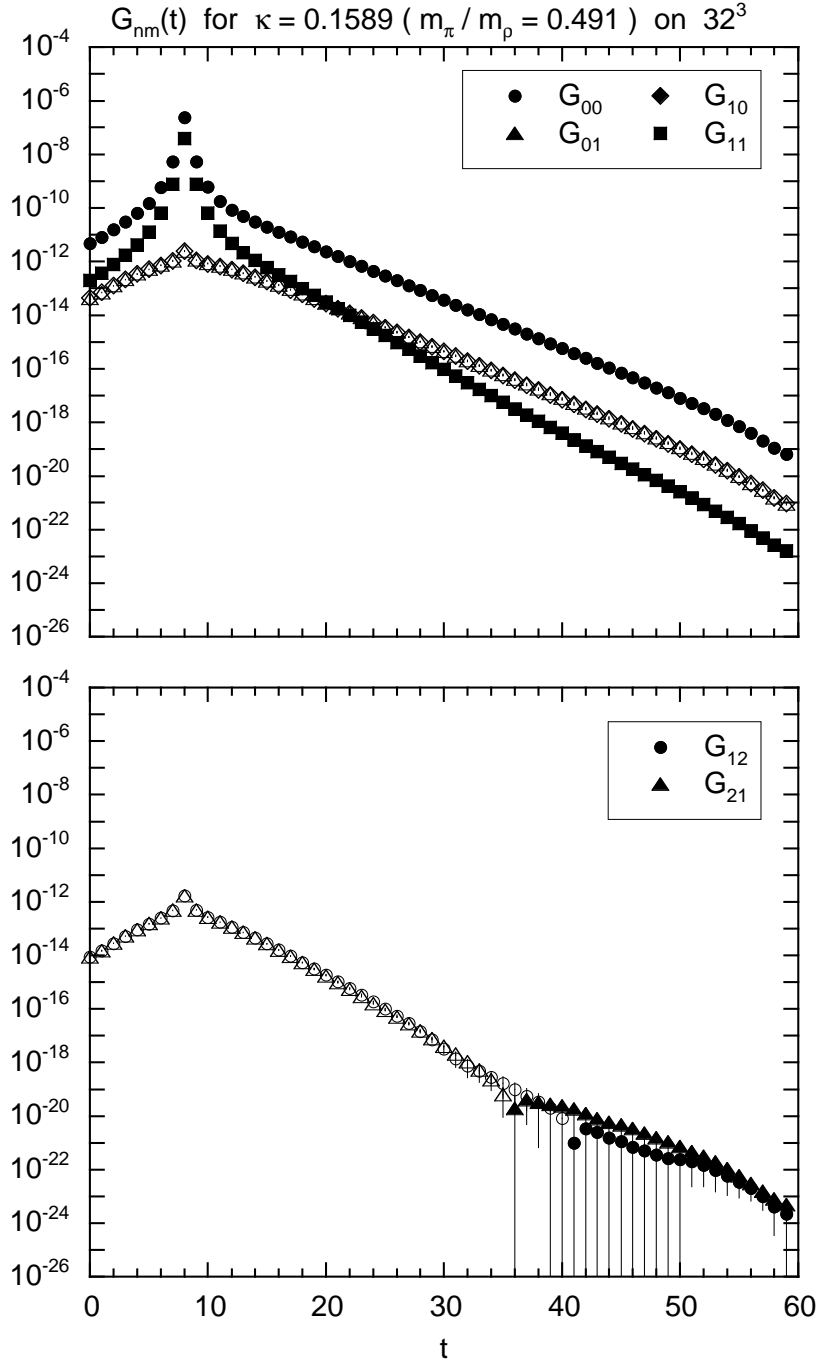, width=11.0cm } }
\vspace{0.5cm}
\caption{ \label{GPK1.fig}
Examples of the pion $4$-point function $G_{nm}(t)$
at $\kappa=0.1589$ ($m_\pi/m_\rho=0.491$) on a $32^3$ lattice.
The two subscripts $n$ and $m$ refer to 
the sink and source momenta $p^2 = (2\pi/L)^2\cdot n$ and
$k^2 = (2\pi/L)^2\cdot m$.
The source is located at $t=8$.
Filled and open symbols indicate positive and negative values.
In the lower frame, large statistical errors are for $G_{12}(t)$,
while those of $G_{21}(t)$ are very small.
}
\end{figure}

The pion $4$-point function $G_{nm}(t)$ defined by (\ref{GPK_I})
is plotted in Fig.~\ref{GPK1.fig} for the same parameter.
The signal is very clear,
and we see that the off-diagonal elements ($n\ne m$) are not negligible.
This means that the overlap is not diagonal,
{\it i.e.} $V_{nm} \not\propto \delta_{nm}$ in (\ref{GPK_III}).
We also observe that the $4$-point function is almost symmetric
under the exchange of the sink and source momenta,
but the statistical errors are not symmetric.
In the lower frame of Fig.~\ref{GPK1.fig}, for example,
$G_{12}(t)$ suffers from large statistical error,
while that of $G_{21}(t)$ is very small.
In the following analysis
we assume symmetry of the magnitude of error, and substitute the component
with large statistical error
by the symmetric partner with smaller error.
We also see evidence of the presence of many exponential terms in
the lower frame of Fig.~\ref{GPK1.fig}.
The sign of $G_{12}(t)$ and $G_{21}(t)$ is flipped at $t \sim 36 - 38$.
This is possible only if more than two exponential terms are present.

In order to examine the effects of diagonalization,
we calculate two ratios defined by
\begin{eqnarray}
&&   R_{n}(t) \equiv G_{nn}(t)
            \cdot \Bigl[ 1 / G_n^\pi (t)             \Bigr]^2
       \label{RaR_def_eq}
\ , \\
&&   D_{n}(t) \equiv \lambda_n (t,t_0)
            \cdot \Bigl[ G_n^\pi (t_0) / G_n^\pi (t) \Bigr]^2
       \label{RaD_def_eq}
\ ,
\end{eqnarray}
where $\lambda_n (t,t_0)$ is the $n$-th eigenvalue of $M(t,t_0)$
calculated with a finite momentum cut-off $p_{\rm cut}^2 = (2\pi/L)^2 \cdot N$.
If the $4$-point function contains only a single exponential term,
{\it i.e.}
$G_{nm}(t) \propto \delta_{nm} \cdot \exp [ - \bar{E}_n \cdot ( t - t_S ) ]$,
then
\begin{equation}
  R_{n}(t) = A \cdot {\rm e}^{ - \Delta E_n \cdot ( t - t_S ) }
\ ,
\end{equation}
where $\Delta E_n \equiv \bar{E}_n - E_n$ and $A$ is a constant.
If the momentum cut-off is sufficiently large,
then the eigenvalue behaves as
$\lambda_n(t,t_0)=\exp [ - \Delta E_n \cdot ( t - t_0 )]$
and
\begin{equation}
  D_n(t) = {\rm e}^{ - \Delta E_n \cdot ( t - t_0 ) }
\ .
\end{equation}
In these cases we can obtain the energy shift $\Delta E_n \equiv \bar{E}_n - E_n$
easily from the ratio $R_n(t)$ or $D_n(t)$ by a single exponential fit.

\begin{figure}
\centerline{ \epsfig{ file=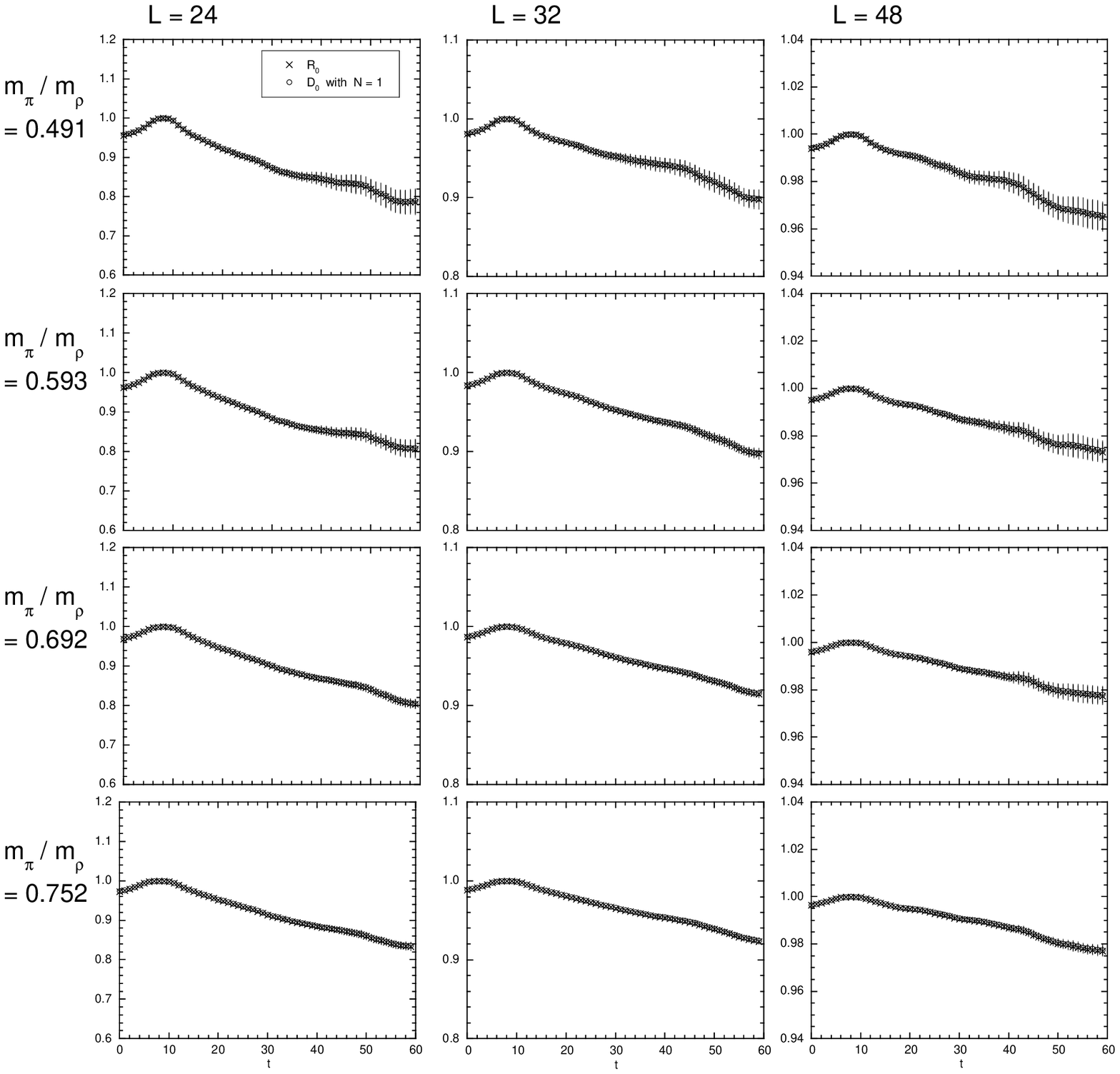, height=18.0cm, width=18.0cm } }
\vspace{0.5cm}
\caption{ \label{DLPA.X.XX.00.fig}
Ratio $R_n(t)$ and $D_n(t)$ for $n=0$ for all quark masses 
and lattice sizes in this work.
Quark mass increases from top to bottom, while lattice size increases 
from left to right.  For diagonalization of $M(t,t_0)$,
the momentum cut-off is set at $N=1$, and
the reference time at $t_0=18$.
}
\end{figure}

In Fig.\ref{DLPA.X.XX.00.fig}
the ratio $R_n(t)$ and $D_n(t)$ for the ground state $n=0$
are plotted for all quark masses and lattice sizes in this work.
For $D_n(t)$ the momentum cut-off $p_{\rm cut}^2 = (2\pi/L)^2 \cdot N$ is set at $N=1$
and the reference time is taken to be $t_0=18$.
We divide $D_n(t)$ by a constant $D_n(t_S)$ to facilitate a comparison with $R_n(t)$.
The statistical errors are very small and the diagonalization does not affect the result.
We also checked the momentum cut-off dependence by taking $N=2$
and confirmed that it is negligible.
In previous calculations of the scattering lengths~\cite{SGK,Kuramashi,GPS,Alford,JLQCD,JLQCD.new,LZCM}
the ratio $R_0(t)$ was used to extract the energy shift $\Delta E_0$.
Our calculation demonstrates the reliability of these calculations.

\begin{figure}
\centerline{ \epsfig{ file=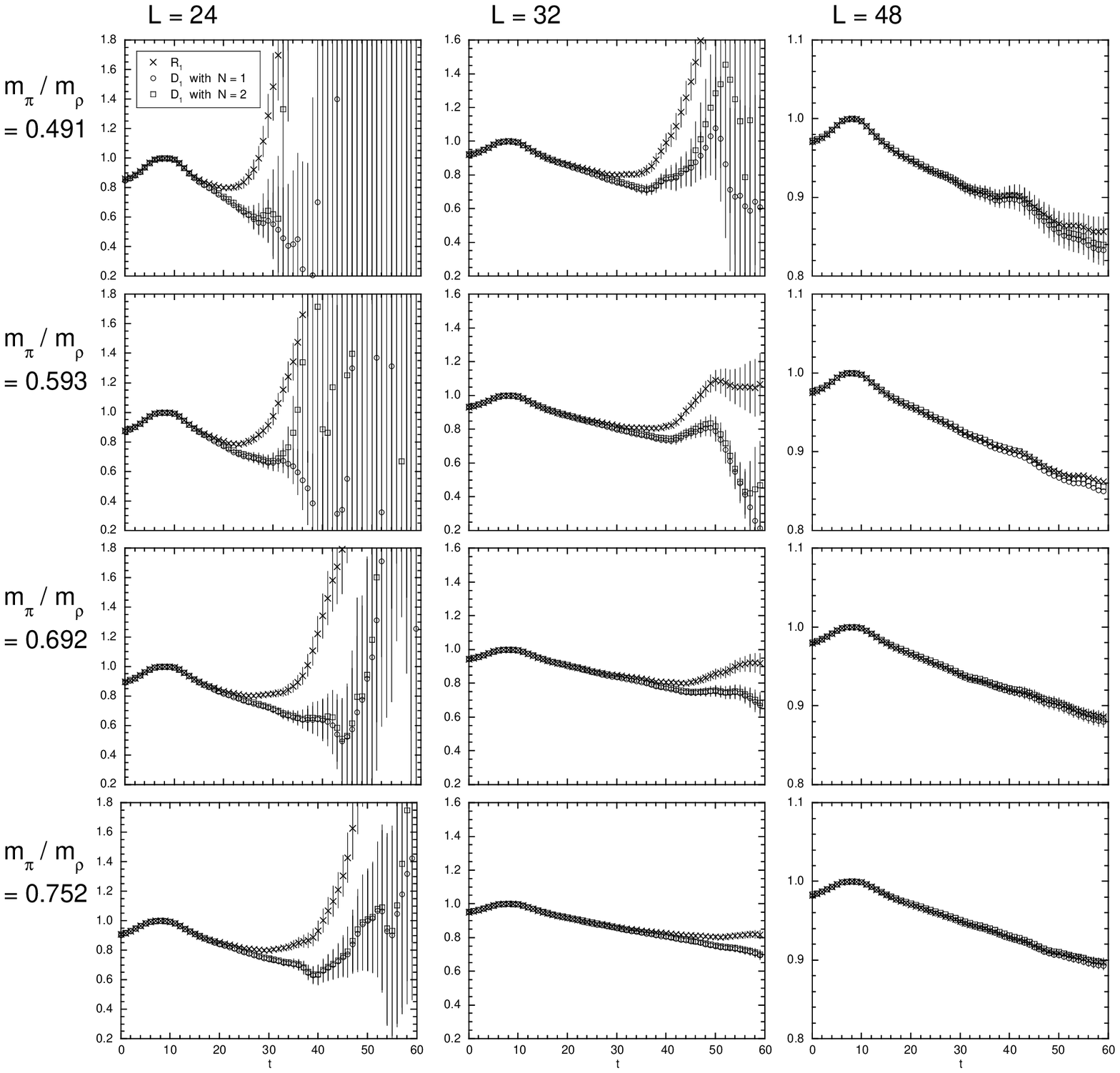, height=18.0cm, width=18.0cm } }
\vspace{0.5cm}
\caption{ \label{DLPA.X.XX.01.fig}
Ratio $R_n(t)$ and $D_n(t)$ for $n=1$ for all quark masses and lattice sizes in this work.
Quark mass increases from top to bottom, while lattice size increases 
from left to right.  For diagonalization of $M(t,t_0)$,
the momentum cut-off is set at $N=1$ and $2$, and
the reference time at $t_0=18$.
}
\end{figure}

We compare the ratios for the first exited state $n=1$
in Fig.~\ref{DLPA.X.XX.01.fig}.
The momentum cut-off is set at $N=1$ and $N=2$.
We divide $D_n(t)$ by a constant $D_n(t_S)$ as for the case of $n=0$.
The diagonalization is effective for smaller quark masses and smaller lattice sizes,
while it is less so for larger quark masses and larger volumes.
The momentum cut-off dependence is negligible for all parameter region, however.
We see a strange behavior near $t = 36$.
We consider that this is either due to insufficient statistics or an effect of
the temporal boundary.
We then fit the ratio by a single exponential form
over the time range consistent with the single exponential behavior.
The fitting range for each parameter
is tabulated in Table~\ref{FResult_n1.table}.

\begin{figure}
\centerline{ \epsfig{ file=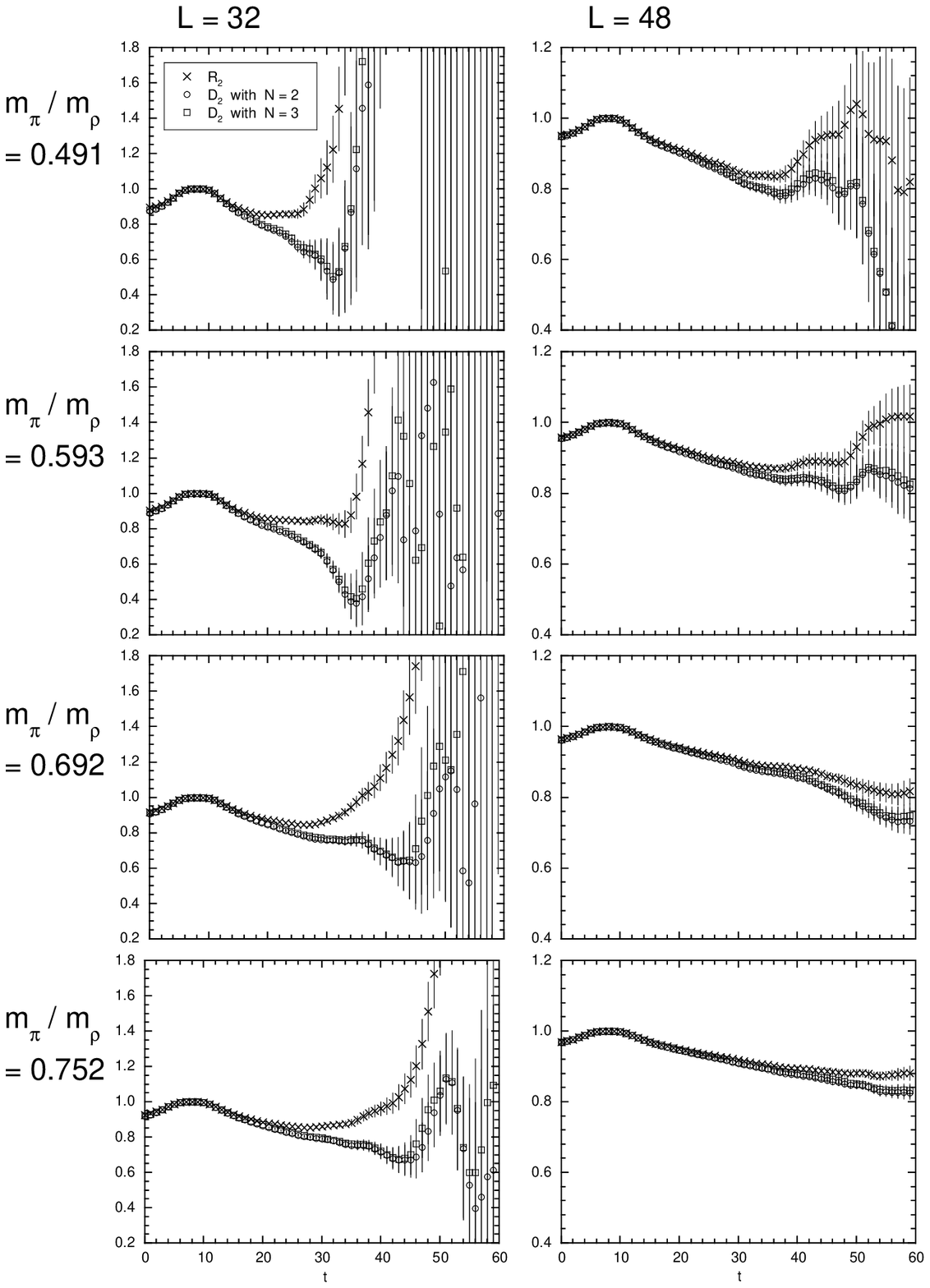, height=18.0cm, width=13.0cm } }
\vspace{0.5cm}
\caption{ \label{DLPA.X.XX.02.fig}
Ratio $R_n(t)$ and $D_n(t)$ for $n=2$ for all quark masses and lattice sizes 
in this work.
Quark mass increases from top to bottom, while lattice size increases  
from left to right. 
For diagonalization of $M(t,t_0)$,
the momentum cut-off is set at $N=2$ and $3$, and
the reference time at $t_0=18$.
}
\end{figure}

A similar comparison for $n=2$ ( the second exited state )
is made in Fig.~\ref{DLPA.X.XX.02.fig}.
The momentum cut-off is set at $N=2$ and $N=3$.
We observe again that the diagonalization is effective for smaller quark masses
and smaller lattice sizes.
The momentum cut-off dependence is small for all parameter region
as for the case of $n=1$.
Compared with the $n=0$ and $n=1$ cases, the signals are noisier.
We observe a strange time dependence in the data
at $m_\pi / m_\rho = 0.491$ and $0.593$ on a $32^3$ lattice
at $t \sim 30 - 46$.
For these data we restrict the fitting range to $t=18-32$.
We remove results at these parameters from our finial analysis.
In other data clear signals of the single exponential behavior are seen
for $t > 18$. The fitting range for each parameter
is listed in Table~\ref{FResult_n2.table}.

From these results we conclude that
the momentum cut-off should be taken $N \geq n$
for the energy shift $\Delta E_n$.
The results of the energy shift $\Delta E_n$
obtained by the single exponential fitting of the ratio $D_n(t)$ are tabulated in
Tables~\ref{FResult_n0.table}, ~\ref{FResult_n1.table}, and ~\ref{FResult_n2.table},
where we take the momentum cut-off $N=n$, and the reference time $t_0 = 18$.
In the tables we also quote
the scattering amplitude $A(\bar{p}_n)$ defined by
\begin{equation}
  A(\bar{p}_n) = \frac{ \tan \delta (\bar{p}_n) }{ \bar{p}_n } \cdot \frac{ \bar{E}_n }{ 2 }
\label{SC-AMP.eq}
\ ,
\end{equation}
where we normalize the amplitude as $\lim_{\bar{p}\to 0} A(\bar{p}) = a_0 \cdot m_\pi$.
%
%
\subsection{ Results of Scattering Length }
For $n=0$ the values of $\bar{p}_n^2$ are very small
as shown in Table~\ref{FResult_n0.table}.  Therefore
we may write $A(\bar{p}_n) / m_\pi^2 \sim a_0 / m_\pi$,
and use results for $n=0$ to evaluate the scattering length.

\begin{figure}
\centerline{ \epsfig{ file=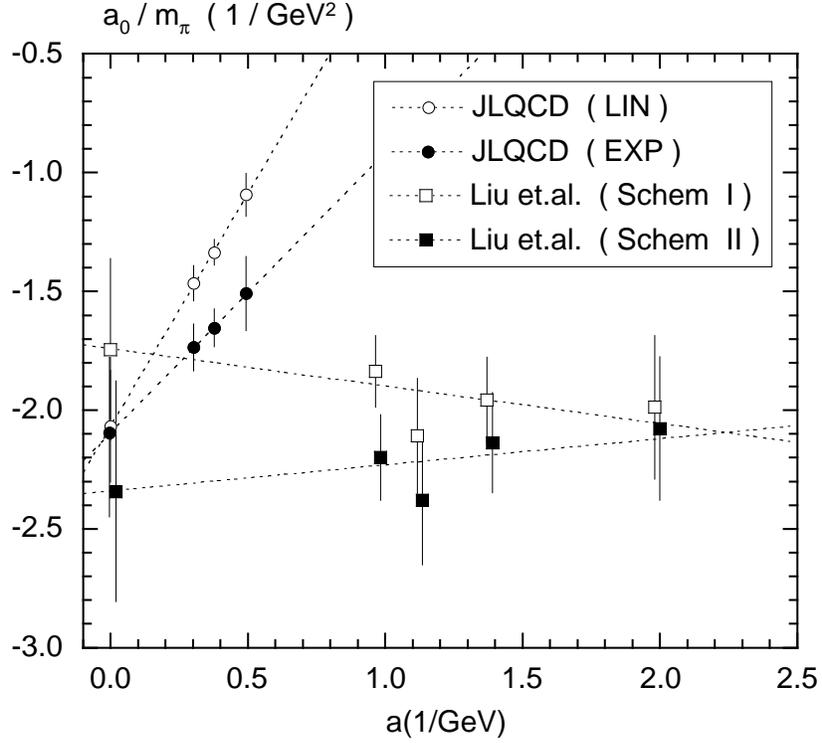, width=11.0cm } }
\vspace{0.5cm}
\caption{ \label{OLD_a0.fig}
Results for scattering length $a_0/m_\pi ({\rm GeV}^2)$
obtained by JLQCD~\protect\cite{JLQCD.new} and by Liu {\it et.al.}~\protect\cite{LZCM}.
}
\end{figure}

In Fig.~\ref{OLD_a0.fig} we recapitulate the recent results of
JLQCD~\cite{JLQCD.new} and Liu {\it et.al.}~\cite{LZCM}
for the $I=2$ pion scattering length.
The two values of Liu {\it et.al.} denoted as (Scheme I) and (Schema II)
refer to their two different treatments of the finite volume corrections.
The two values of JLQCD correspond to
two different fitting functions for extraction of the energy shift
from the ratio $R_0(t)$, (LIN) used a linear fit in $t$
while (EXP) employs a single exponential in $t$.
Figure~\ref{OLD_a0.fig} shows that the lattice cut-off effect
is strongly dependent on the choice of the fitting function.
However, the dependence disappears toward the continuum limit.
Compared with JLQCD the lattice cut-off effect of Liu {\it et.al.}
is very small, since their calculation is carried out
with an improved gauge and improved Wilson fermion action on anisotropic lattices,
while the actions of JLQCD are the standard plaquette and the Wilson fermion actions.
The values extrapolated to the continuum limit are consistent with
the CHPT prediction~\cite{CHPT_a0} as shown in Table~\ref{OLD_a0.table}.

\begin{figure}
\centerline{ \epsfig{ file=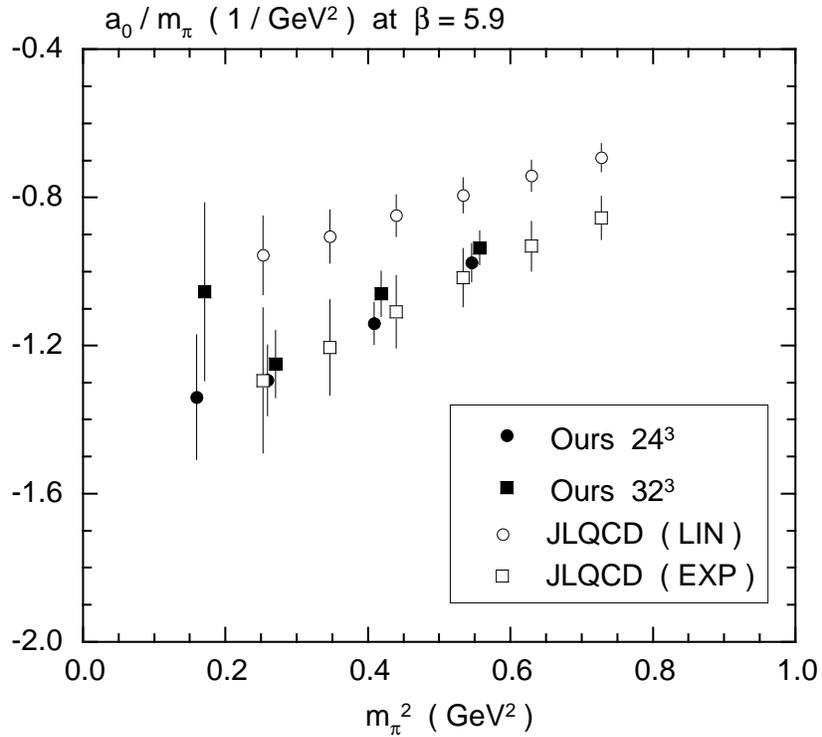, width=11.0cm } }
\vspace{0.5cm}
\caption{ \label{COMP.JLQCD_a0.fig}
Comparison of our results on $24^3$ and $32^3$ lattices with those of 
JLQCD on a $16^3$ lattice at $\beta = 5.9$~\protect\cite{JLQCD.new}.
}
\end{figure}

Since we use the same actions as those of JLQCD,
we compare our results with theirs at the same gauge coupling constant
$\beta=5.9$ in Fig.~\ref{COMP.JLQCD_a0.fig}.
Here our data on a $48^3$ lattice are omitted,
because those are consistent with the results on $24^3$ and $32^3$ lattices
within very large statistical errors of those on the $48^3$ lattice
(see Table~\ref{FResult_n0.table}).
Our data for the scattering length are different from those of JLQCD
obtained by a linear fit (LIN) by about $2.5\sigma$,
whereas we find consistency among results obtained with the exponential fitting
for four different lattice sizes,
{\it i.e.}
$24^3$, $32^3$, $48^3$ from the present work, and $16^3$ from JLQCD.
In Fig.~\ref{COMP.JLQCD_a0.fig}
we observe that both ours and JLQCD results at $\beta=5.9$
are far from the CHPT prediction $a_0/m_\pi = - 2.265(51) ~1/{\rm GeV}^2$.
This is due to finite lattice cut-off effects, which are rather large for the
standard actions as shown in Fig.~\ref{OLD_a0.fig}.

Here we comment on the choice of the fitting function for the ratio $R_0(t)$.
In our analysis we have assumed a single exponential behavior,
{\it i.e.} $R_0(t) \sim Z \cdot \exp( -\Delta E_0 (t-t_S) )$ for large $t - t_S$.
The validity of this assumption was partially
examined by Sharpe {\it et.al.}~\cite{SGK}.
Writing
\begin{equation}
  R_0(t) = Z \cdot \biggl(
               1
             - \Delta E_0                  \cdot ( t - t_S )
             + \frac{1}{2} \cdot ( \Delta E_0' )^2 \cdot ( t - t_S )^2
             + {\rm O}( (t-t_S)^3 )
          \biggr)
\ ,
\label{GB_eq}
\end{equation}
they showed in time-ordered perturbation theory that
the lattice value of $\Delta E_0 $ is related to the scattering length
by the L\"uscher relation (\ref{LUSCHER_eq}) up to corrections of $O(L^{-5})$.
By a similar calculation, one easily shows that the value of $\Delta E_0'$
deviates from $\Delta E_0 $ by terms of $O(L^{-5})$.
These effects occur due to intermediate off-shell two-pion states.

In the context of our analysis,
the momentum cut-off dependence is negligible as discussed in Sec.~\ref{FOUR_S}.
This means that the effects
due to the intermediate off-shell two-pion states are negligible.
Thus the correction of $O(L^{-5})$ for $\Delta E_0$ and $\Delta E_0'$ is sufficiently small,
and the time behavior can be regarded as a single exponential function in our simulation.

To check this point more explicitly,
we calculate the scattering length
with the energy shift obtained with both the linear
and the single exponential function in $t$ as was done by JLQCD.
Results are tabulated in Table~\ref{SCL.table},
which shows that the two set of values are consistent within statistical errors,
and have no volume dependence.
These facts indicate that the deviation of JLQCD results
between the two fitting functions
comes from the approximation of the exponential function to the linear function in $t$,
{\it i.e.}
the value of $\Delta E_0 \cdot (t-t_S) \sim 1/L^3 \cdot (t-t_S)$ is not
small enough to justify such an approximation due to small lattice sizes.

Another comment concerns the quenching effect on the ratio $R_0(t)$.
Bernard and Golterman derived the same time behavior (\ref{GB_eq})
using quenched chiral perturbation theory (qCHPT)~\cite{Bernard-Golterman}.
They predicted that the scattering length obtained with quenched approximation
is divergent in the chiral limit as $a_0 \sim 1/m_\pi$.
These effects are attributed to non-unitarity of the quenched theory.
The same results are also obtained by
Colangelo and Pallante~\cite{Colangelo-Pallante}.
Divergence in scattering lengths in the chiral limit can also occur
if one uses chirally non-symmetric lattice fermion action,
for example the Wilson fermion action.

In Fig.~\ref{COMP.JLQCD_a0.fig}
we do not observe signs of divergence toward the chiral limit.
We consider that the effects of quenching and broken chiral symmetry
are still too small to affect data at our simulation points.

The quenching problems can also occur for non-zero momenta,
{\it i.e.}
it is not proven that the pion $4$-point function $G_{nm}(t)$ behave as
a multi-exponential function in $t$ as (\ref{GPK_III}) and
the diagonalization method can be used.
In this work we assume that such effects are small in our simulation points
as confirmed for the zero momentum case.
Investigation of the quenching effects for the scattering length and the phase shift
by lattice simulations
with small quark masses is an important future work.
%
%
\subsection{ Results of scattering phase shift }
The energy shift $\Delta E_n \equiv \bar{E}_n - E_n$ and
the phase shift $\delta(\bar{p}_n)$
at our simulation points are tabulated
in Tables~\ref{FResult_n0.table}, \ref{FResult_n1.table}, and \ref{FResult_n2.table}.
The scattering amplitude $A(\bar{p}_n)$ defined by (\ref{SC-AMP.eq}) are also
included in these tables.

\begin{figure}
\centerline{ \epsfig{ file=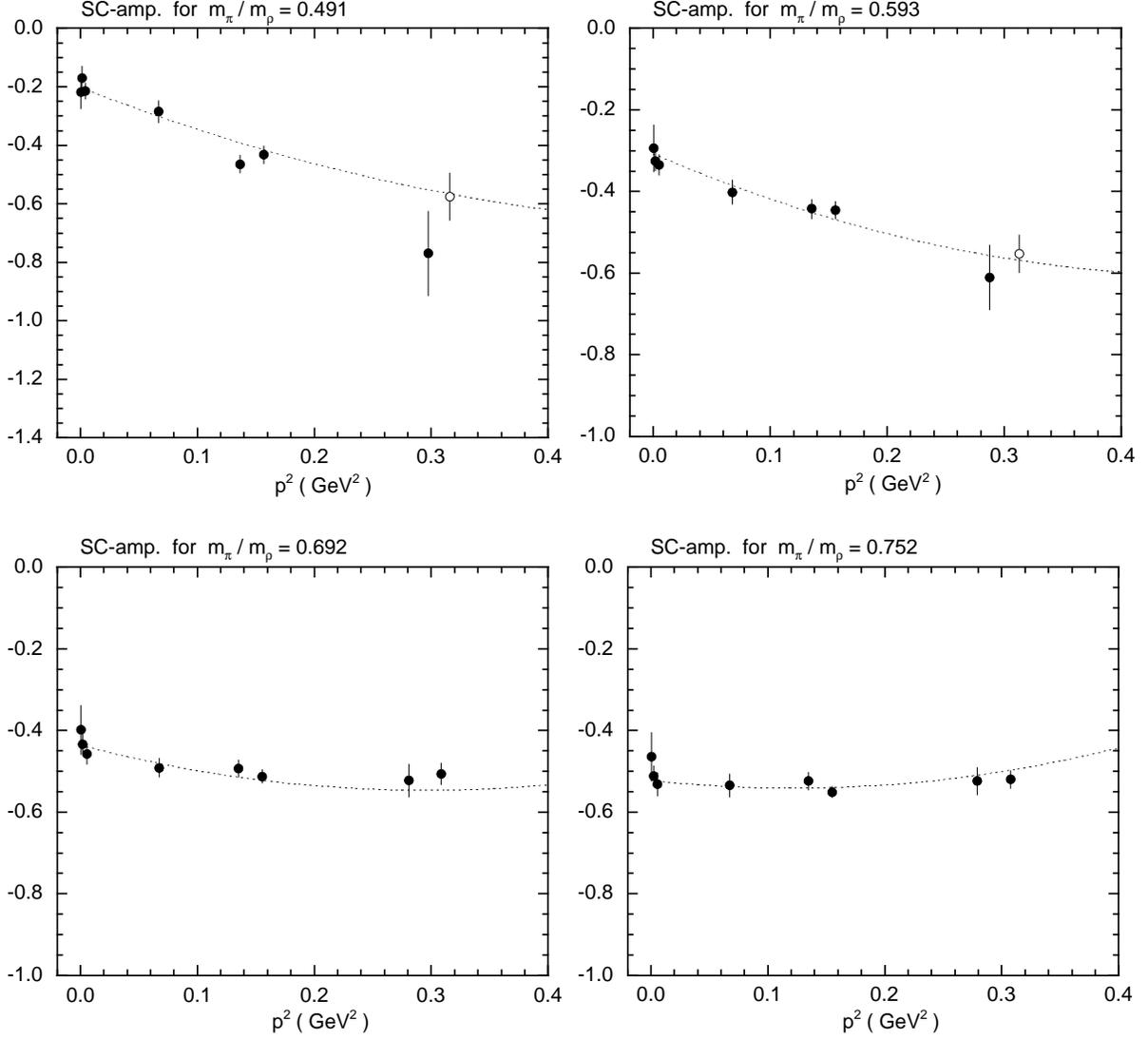, width=16.0cm } }
\vspace{0.5cm}
\caption{ \label{AMP.fig}
Scattering amplitude
$A(\bar{p}) = \tan \delta (\bar{p}) / \bar{p} \cdot \bar{E} / 2$
for fixed quark masses.
The fit curve is also plotted.
The open symbols indicate data omitted in the fitting procedure.
}
\end{figure}

In Fig.~\ref{AMP.fig} we plot the amplitude at fixed quark mass as a function
of the momentum $\bar{p}_n^2$.
In order to obtain the scattering phase shift
for various momenta at the physical pion mass,
we extrapolate our data with the following fitting assumption :
\begin{eqnarray}
&& A(\bar{p} ) \equiv \frac{ \tan \delta(\bar{p}) }{ \bar{p} } \cdot \frac{ \bar{E} }{ 2 }   \cr
&& \qquad
       =
         A_{00}
       + A_{10} \cdot (m_\pi^2)
       + A_{20} \cdot (m_\pi^2)^2
       + A_{01} \cdot (\bar{p}^2)     \cr
&& \qquad \quad
       + A_{11} \cdot (m_\pi^2 ) (\bar{p}^2)
       + A_{02} \cdot (\bar{p}^2)^2
\label{FTF-SC_amp.eq}
\ .
\end{eqnarray}
Here $A_{10}$ corresponds to $a_0 / m_\pi$.
In Fig.~\ref{AMP.fig} we omit data plotted with open symbols in the fitting.
They are for the momentum $n=2$ on a $32^3$ lattice
at $m_\pi/m_\rho = 0.491$ and $0.593$
for which a clear plateau in $D_n(t)$ is absent.
It should be noted that the constant term $A_{00}$ vanishes
if the effects of quenching and chiral symmetry breaking are negligible.
We tried to fit our data both with and without the assumption $A_{00}=0$.
The results, tabulated in Table~\ref{A_fit.table},
show that the latter fit yields a value of $A_{00}$ which is 1.7$\sigma$
away from zero.
The other parameters, such as $A_{10}$
which are physically more relevant,
are consistent between the two types of fits, however.
From these observations we adopt the value with the assumption of $A_{00}=0$.
The fit curves for this fitting are also plotted in Fig.~\ref{AMP.fig}.

\begin{figure}
\centerline{ \epsfig{ file=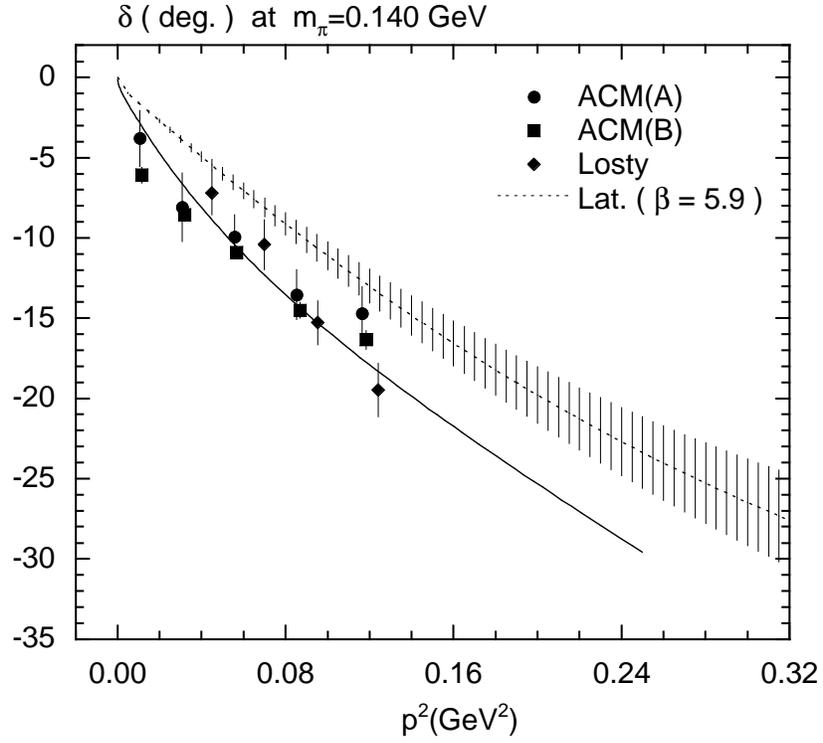, width=11.0cm } }
\vspace{0.5cm}
\caption{ \label{DCOM_EXPT.fig}
Comparison of our results for scattering phase shift $\delta(p)$
at physical pion mass with experiments~\protect\cite{ACM_expt,Losty_expt}.
}
\end{figure}

We present our results for the phase shift $\delta(p)$
at the physical pion mass obtained with the fitting (\ref{FTF-SC_amp.eq})
with the assumption $A_{00}=0$ in Fig.~\ref{DCOM_EXPT.fig}.
The filled points are experimental results~\cite{ACM_expt,Losty_expt}.
The values of the phase shift at several momenta
are tabulated in Table~\ref{D_LAT.table}.
Our results are 30\% smaller in magnitude than the experiments.
A possible origin of the discrepancy is finite lattice spacing effects.
As we saw in Fig.~\ref{OLD_a0.fig} the JLQCD results for scattering length
show a sizable scaling violation.
Hence that of the scattering phase shift cannot be considered small.
Further calculations nearer to the continuum limit or
calculations with improved actions
are desirable to obtain the continuum result of the phase shift.
%
%
\section{\label{FIVE_S} Conclusions }
We have shown in this work that calculations of the scattering length are possible
with present computing resources.
The quenched approximation we employed has theoretical issues
regarding the chiral extrapolation.  We see no problem, either theoretically or
computationally, in avoiding this problem by going to full QCD calculations,
for the simplest case of the $I=2$ two-pion system.
The cases of $I=0$ and $I=1$, which are richer in physics content,
are much more difficult
from the computational point of view.  Algorithmic advances are presumably
needed to
evaluate the box and two-loop diagrams with good precision for non-zero momenta,
which are needed to extract the two-pion energy eigenvalues in these channels.

Another implication of this work is feasibility of a direct calculation of
the $K\to\pi\pi$ decay amplitude using the method of Lellouch and L\"uscher.
Diagonalization of the pion 4-point function yields the two-pion eigenstate
for non-zero relative momenta, which can be used as the final state for the
$K\to\pi\pi$ Green function needed in their method.  Executing this program
for the $I=2$ channel would be an interesting step to take to solve
this long-standing problem.
%
%
\section*{Acknowledgment}
This work is supported in part by Grants-in-Aid of the Ministry of Education
(Nos.
11640294, 
12304011, 
12640253, 
12740133, 
13640259, 
13640260, 
13135204, 
14046202, 
14740173, 
).
VL is supported by the Research for Future Program of JSPS
(No. JSPS-RFTF 97P01102).
Simulations were performed on the parallel computer CP-PACS.
%
%

%
%
\newpage
%
%
%
%
\newpage
%
%
\newpage
%
%
\newpage
%
%
\newpage
%
%
\newpage
%
%
%
%
\newpage
%
%
\newpage
%
%
\begin{table}[t]
\begin{center}
\begin{tabular}{lrrr}
%
 $j$ &  $n=0$ ($N_n=1$)   & $n=1$ ($N_n=6$)   &   $n=2$ ($N_n=12$) \\
\hline
 1   &  $-8.913632922$    & $-1.211335686$    &   $-5.096565798$  \\
 2   &  $16.532315957$    & $23.243221879$    &   $25.661192388$  \\
 3   &  $ 8.401923974$    & $13.059376755$    &   $ 4.254135936$  \\
 4   &  $ 6.945807927$    & $13.731214368$    &   $14.867522887$  \\
 5   &  $ 6.426119102$    & $11.308518083$    &   $ 2.283549584$  \\
 6   &  $ 6.202149045$    & $13.140942288$    &   $14.148854520$  \\
 7   &  $ 6.098184125$    & $11.067054131$    &   $ 2.051601110$  \\
 8   &  $ 6.048263469$    & $13.032596991$    &   $14.031382623$  \\
 9   &  $ 6.023881707$    & $11.016034293$    &   $ 2.011078709$  \\
10   &  $ 6.011862830$    & $13.007939537$    &   $14.007265604$  \\
\end{tabular}
\end{center}
\caption{\label{ZETA.table}
Values of the zeta function $Z_{00}( j ; n )$ and $N_n=\sum_{{\bf m}\in {\bf Z}^3} \delta( m^2 - n )$
for momenta $p^2=(2\pi/L)^2\cdot n$.
}
\end{table}
%
%
\newpage
\begin{table}[t]
\begin{center}
\begin{tabular}{ll rrrr}
%
 &                                               & $\kappa=0.1589$ & $\kappa=0.1583$ & $\kappa=0.1574$ & $\kappa=0.1566$ \\
 & \multicolumn{1}{r}{$m_\pi/m_\rho$}            & $0.491(2)     $ & $0.593(1)     $ & $0.692(1)     $ & $0.752(1)     $ \\
 & \multicolumn{1}{r}{$m_\pi^2$ $({\rm GeV}^2)$} & $0.16113(97)  $ & $0.26026(90)  $ & $0.40896(91)  $ & $0.5468(11)   $ \\
\hline
$V=24^3$        \\
\multicolumn{2}{l}{Fitting Range}                          & $18-40     $ & $18-40     $ & $18-44     $ & $18-44     $ \\
$\Delta E_n            $ & $(\times 10^{-4}\ {\rm GeV}  )$ & $98(14)    $ & $97.9(83)  $ & $86.3(53)  $ & $75.5(47)  $ \\
$\bar{p}_n^2 - p_n^2   $ & $(\times 10^{-4}\ {\rm GeV}^2)$ & $39.5(56)  $ & $50.1(43)  $ & $55.4(35)  $ & $55.9(35)  $ \\
$\bar{n}-n             $ & $(\times 10^{-3}             )$ & $15.4(22)  $ & $19.5(17)  $ & $21.6(13)  $ & $21.8(14)  $ \\
$\bar{p}_n^2           $ & $(\times 10^{-4}\ {\rm GeV}^2)$ & $39.5(56)  $ & $50.1(43)  $ & $55.4(35)  $ & $55.9(35)  $ \\
$\delta(\bar{p}_n)     $ & $({\rm deg.}                 )$ & $-1.91(37) $ & $-2.64(31) $ & $-3.03(25) $ & $-3.07(26) $ \\
$A(\bar{p}_n)          $ &                                 & $-0.214(27)$ & $-0.335(25)$ & $-0.458(24)$ & $-0.532(29)$ \\
$A(\bar{p}_n) / m_\pi^2$ & $(1/{\rm GeV}^2)              $ & $-1.34(17) $ & $-1.293(96)$ & $-1.119(59)$ & $-0.975(51)$ \\
\hline
\hline
%
%
$V=32^3$        \\
\multicolumn{2}{l}{Fitting Range}                          & $18-44      $ & $18-44      $ & $18-44      $ & $18-44     $ \\
$\Delta E_n            $ & $(\times 10^{-4}\ {\rm GeV}  )$ & $31.3(77)   $ & $38.6(31)   $ & $33.0(21)   $ & $29.3(16)  $ \\
$\bar{p}_n^2 - p_n^2   $ & $(\times 10^{-4}\ {\rm GeV}^2)$ & $12.6(31)   $ & $19.7(16)   $ & $21.1(14)   $ & $21.7(12)  $ \\
$\bar{n}-n             $ & $(\times 10^{-3}             )$ & $8.7(22)    $ & $13.7(11)   $ & $14.66(95)  $ & $15.02(82) $ \\
$\bar{p}_n^2           $ & $(\times 10^{-4}\ {\rm GeV}^2)$ & $12.6(31)   $ & $19.7(16)   $ & $21.1(14)   $ & $21.7(12)  $ \\
$\delta(\bar{p}_n)     $ & $({\rm deg.}                 )$ & $-0.86(31)  $ & $-1.62(18)  $ & $-1.78(16)  $ & $-1.84(14) $ \\
$A(\bar{p}_n)          $ &                                 & $-0.170(39) $ & $-0.325(24) $ & $-0.433(25) $ & $-0.512(25)$ \\
$A(\bar{p}_n) / m_\pi^2$ & $(1/{\rm GeV}^2              )$ & $-1.05(24)  $ & $-1.250(91) $ & $-1.060(61) $ & $-0.936(46)$ \\
\hline
\hline
%
%
$V=48^3$        \\
\multicolumn{2}{l}{Fitting Range}                          & $18-44     $ & $18-44     $ & $18-4      $ & $18-44     $ \\
$\Delta E_n            $ & $(\times 10^{-4}\ {\rm GeV}  )$ & $11.8(33)  $ & $10.0(21)  $ & $8.6(14)   $ & $7.5(10)   $ \\
$\bar{p}_n^2 - p_n^2   $ & $(\times 10^{-4}\ {\rm GeV}^2)$ & $4.7(13)   $ & $5.1(11)   $ & $5.51(89)  $ & $5.55(76)  $ \\
$\bar{n}-n             $ & $(\times 10^{-3}             )$ & $7.4(21)   $ & $7.9(17)   $ & $8.6(14)   $ & $8.7(12)   $ \\
$\bar{p}_n^2           $ & $(\times 10^{-4}\ {\rm GeV}^2)$ & $4.7(13)   $ & $5.1(11)   $ & $5.51(89)  $ & $5.55(76)  $ \\
$\delta(\bar{p}_n)     $ & $({\rm deg.}                 )$ & $-0.68(27) $ & $-0.74(23) $ & $-0.84(20) $ & $-0.85(17) $ \\
$A(\bar{p}_n)          $ &                                 & $-0.217(57)$ & $-0.294(58)$ & $-0.399(60)$ & $-0.464(59)$ \\
$A(\bar{p}_n) / m_\pi^2$ & $(1/{\rm GeV}^2              )$ & $-1.35(36) $ & $-1.13(22) $ & $-0.98(15) $ & $-0.85(11) $ \\
\end{tabular}
\end{center}
\caption{\label{FResult_n0.table}
Results for $n=0$ with momentum cut-off $N=0$ and $t_0=18$.
The scattering amplitude $A(\bar{p}_n)$ is defined by
$A(\bar{p}_n) = \tan \delta ( \bar{p}_n ) / \bar{p}_n \cdot \bar{E}_n / 2$.
$A(\bar{p}_n) / m_\pi^2$ corresponds to $a_0 / m_\pi$.
}
\end{table}
%
%
\newpage
\begin{table}[t]
\begin{center}
\begin{tabular}{ll rrrr}
%
 &                                               & $\kappa=0.1589$ & $\kappa=0.1583$ & $\kappa=0.1574$ & $\kappa=0.1566$ \\
 & \multicolumn{1}{r}{$m_\pi/m_\rho$}            & $0.491(2)     $ & $0.593(1)     $ & $0.692(1)     $ & $0.752(1)     $ \\
 & \multicolumn{1}{r}{$m_\pi^2$ $({\rm GeV}^2)$} & $0.16113(97)  $ & $0.26026(90)  $ & $0.40896(91)  $ & $0.5468(11)   $ \\
\hline
$V=24^3$        \\
\multicolumn{2}{l}{Fitting Range}                       & $18-32    $ & $18-32     $ & $18-44     $ & $18-44     $ \\
$\Delta E_n         $ & $(\times 10^{-3}\ {\rm GeV}  )$ & $63(11)   $ & $42.7(54)  $ & $29.8(22)  $ & $25.1(16)  $ \\
$\bar{p}_n^2 - p_n^2$ & $(\times 10^{-3}\ {\rm GeV}^2)$ & $41.3(73) $ & $31.3(39)  $ & $24.5(18)  $ & $22.5(14)  $ \\
$\bar{n}-n          $ & $(\times 10^{-2}             )$ & $16.1(28) $ & $12.2(15)  $ & $9.54(71)  $ & $8.79(56)  $ \\
$\bar{p}_n^2        $ & $(\times 10^{-2}\ {\rm GeV}^2)$ & $29.77(73)$ & $28.76(39) $ & $28.08(18) $ & $27.89(14) $ \\
$\delta(\bar{p}_n)  $ & $({\rm deg.}                 )$ & $-31.8(57)$ & $-23.8(30) $ & $-18.5(14) $ & $-17.0(11) $ \\
$A(\bar{p}_n)       $ &                                 & $-0.77(14)$ & $-0.611(79)$ & $-0.523(40)$ & $-0.524(34)$ \\
\hline
\hline
%
%
$V=32^3$        \\
\multicolumn{2}{l}{Fitting Range}                       & $18-36     $ & $18-40     $ & $18-44     $ & $18-44     $ \\
$\Delta E_n         $ & $(\times 10^{-3}\ {\rm GeV}  )$ & $22.5(15)  $ & $17.89(81) $ & $15.11(50) $ & $13.05(32) $ \\
$\bar{p}_n^2 - p_n^2$ & $(\times 10^{-3}\ {\rm GeV}^2)$ & $12.58(85) $ & $11.42(52) $ & $11.27(35) $ & $10.88(26) $ \\
$\bar{n}-n          $ & $(\times 10^{-2}             )$ & $8.72(59)  $ & $7.92(36)  $ & $7.82(24)  $ & $7.54(18)  $ \\
$\bar{p}_n^2        $ & $(\times 10^{-2}\ {\rm GeV}^2)$ & $15.678(85)$ & $15.562(52)$ & $15.548(35)$ & $15.508(26)$ \\
$\delta(\bar{p}_n)  $ & $({\rm deg.}                 )$ & $-16.9(12) $ & $-15.29(71)$ & $-15.08(47)$ & $-14.54(36)$ \\
$A(\bar{p}_n)       $ &                                 & $-0.432(30)$ & $-0.445(21)$ & $-0.513(16)$ & $-0.551(14)$ \\
\hline
\hline
%
%
$V=48^3$        \\
\multicolumn{2}{l}{Fitting Range}                       & $18-44     $ & $18-44     $ & $18-44     $ & $18-44     $ \\
$\Delta E_n         $ & $(\times 10^{-3}\ {\rm GeV}  )$ & $6.24(81)  $ & $6.08(43)  $ & $5.10(24)  $ & $4.31(22)  $ \\
$\bar{p}_n^2 - p_n^2$ & $(\times 10^{-3}\ {\rm GeV}^2)$ & $2.96(39)  $ & $3.46(25)  $ & $3.51(16)  $ & $3.37(18)  $ \\
$\bar{n}-n          $ & $(\times 10^{-2}             )$ & $4.62(60)  $ & $5.40(39)  $ & $5.48(25)  $ & $5.25(27)  $ \\
$\bar{p}_n^2        $ & $(\times 10^{-2}\ {\rm GeV}^2)$ & $6.705(39) $ & $6.755(25) $ & $6.760(16) $ & $6.746(18) $ \\
$\delta(\bar{p}_n)  $ & $({\rm deg.}                 )$ & $-8.8(12)  $ & $-10.35(76)$ & $-10.50(49)$ & $-10.06(53)$ \\
$A(\bar{p}_n)       $ &                                 & $-0.285(38)$ & $-0.402(29)$ & $-0.492(23)$ & $-0.535(28)$ \\
\end{tabular}
\end{center}
\caption{\label{FResult_n1.table}
Results for $n=1$ with the momentum cut-off $N=1$ and $t_0=18$.
The scattering amplitude $A(\bar{p}_n)$ is defined by
$A(\bar{p}_n) = \tan \delta ( \bar{p}_n ) / \bar{p}_n \cdot \bar{E}_n / 2$.
}
\end{table}
%
%
%
\newpage
\begin{table}[t]
\begin{center}
\begin{tabular}{ll rrrr}
%
 &                                               & $\kappa=0.1589$ & $\kappa=0.1583$ & $\kappa=0.1574$ & $\kappa=0.1566$ \\
 & \multicolumn{1}{r}{$m_\pi/m_\rho$}            & $0.491(2)     $ & $0.593(1)     $ & $0.692(1)     $ & $0.752(1)     $ \\
 & \multicolumn{1}{r}{$m_\pi^2$ $({\rm GeV}^2)$} & $0.16113(97)  $ & $0.26026(90)  $ & $0.40896(91)  $ & $0.5468(11)   $ \\
\hline
$V=32^3$        \\
\multicolumn{2}{l}{Fitting Range}                       & $18-32     $ & $18-32     $ & $18-40     $ & $18-44     $ \\
$\Delta E_n         $ & $(\times 10^{-3}\ {\rm GeV}  )$ & $40.9(56)  $ & $32.9(27)  $ & $24.3(12)  $ & $20.93(86) $ \\
$\bar{p}_n^2 - p_n^2$ & $(\times 10^{-3}\ {\rm GeV}^2)$ & $27.8(38)  $ & $24.5(20)  $ & $20.3(10)  $ & $19.15(80) $ \\
$\bar{n}-n          $ & $(\times 10^{-2}             )$ & $19.3(27)  $ & $17.0(14)  $ & $14.10(72) $ & $13.28(55) $ \\
$\bar{p}_n^2        $ & $(\times 10^{-2}\ {\rm GeV}^2)$ & $31.62(38) $ & $31.30(20) $ & $30.87(10) $ & $30.756(79)$ \\
$\delta(\bar{p}_n)  $ & $({\rm deg.}                 )$ & $-25.2(34) $ & $-22.3(18) $ & $-18.47(93)$ & $-17.41(72)$ \\
$A(\bar{p}_n)       $ &                                 & $-0.576(81)$ & $-0.552(46)$ & $-0.507(26)$ & $-0.520(22)$ \\
\hline
\hline
%
%
$V=48^3$        \\
\multicolumn{2}{l}{Fitting Range}                       & $18-36     $ & $18-44     $ & $18-44     $ & $18-44     $ \\
$\Delta E_n         $ & $(\times 10^{-3}\ {\rm GeV}  )$ & $15.9(10)  $ & $11.40(61) $ & $9.20(38)  $ & $7.81(33)  $ \\
$\bar{p}_n^2 - p_n^2$ & $(\times 10^{-3}\ {\rm GeV}^2)$ & $8.57(55)  $ & $7.11(38)  $ & $6.75(28)  $ & $6.41(27)  $ \\
$\bar{n}-n          $ & $(\times 10^{-2}             )$ & $13.37(86) $ & $11.09(59) $ & $10.53(44) $ & $10.00(42) $ \\
$\bar{p}_n^2        $ & $(\times 10^{-2}\ {\rm GeV}^2)$ & $13.675(55)$ & $13.529(38)$ & $13.493(28)$ & $13.459(27)$ \\
$\delta(\bar{p}_n)  $ & $({\rm deg.}                 )$ & $-17.5(11) $ & $-14.56(77)$ & $-13.83(58)$ & $-13.15(55)$ \\
$A(\bar{p}_n)       $ &                                 & $-0.464(31)$ & $-0.442(24)$ & $-0.493(21)$ & $-0.524(22)$ \\
\end{tabular}
\end{center}
\caption{\label{FResult_n2.table}
Results for $n=2$ with the momentum cut-off $N=2$ and $t_0=18$.
The scattering amplitude $A(\bar{p}_n)$ is defined by
$A(\bar{p}_n) = \tan \delta ( \bar{p}_n ) / \bar{p}_n \cdot \bar{E}_n / 2$.
}
\end{table}
%
%
\newpage
\begin{table}[t]
\begin{center}
\begin{tabular}{lll}
                               & $a_0 / m_\pi$ $(1/{\rm GeV}^2)$ & $a_0 \cdot m_\pi$  \\
\hline
JLQCD ( LIN )                  & $-2.07(24)$                     & $-0.0406(47)$ \\
JLQCD ( EXP )                  & $-2.09(35)$                     & $-0.0410(69)$ \\
Liu {\it et.al.} ( Scheme I  ) & $-1.75(38)$                     & $-0.0342(75)$ \\
Liu {\it et.al.} ( Scheme II ) & $-2.34(46)$                     & $-0.0459(91)$ \\
CHPT                           & $-2.265(51)$                    & $-0.0444(10)$ \\
\end{tabular}
\end{center}
\caption{\label{OLD_a0.table}
Recent results for the scattering length $a_0$ in the continuum limit.
CHPT refers to the prediction of chiral perturbation theory. The error 
for this case shows theoretical uncertainties.
}
\end{table}
%
%
\begin{table}[t]
\begin{center}
\begin{tabular}{ll rrrr}
 &                                               & $\kappa=0.1589$ & $\kappa=0.1583$ & $\kappa=0.1574$ & $\kappa=0.1566$ \\
 & \multicolumn{1}{r}{$m_\pi/m_\rho$}            & $0.491(2)     $ & $0.593(1)     $ & $0.692(1)     $ & $0.752(1)     $ \\
 & \multicolumn{1}{r}{$m_\pi^2$ $({\rm GeV}^2)$} & $0.16113(97)  $ & $0.26026(90)  $ & $0.40896(91)  $ & $0.5468(11)   $ \\
\hline
$24^3$
& LIN  & $-1.23(14) $ & $-1.194(82)$ & $-1.042(51)$ & $-0.917(46)$  \\
& EXP  & $-1.34(17) $ & $-1.293(96)$ & $-1.119(59)$ & $-0.975(51)$  \\
\hline
$32^3$
& LIN  & $-1.02(23) $ & $-1.207(85)$ & $-1.029(58)$ & $-0.912(43)$  \\
& EXP  & $-1.05(24) $ & $-1.250(91)$ & $-1.060(61)$ & $-0.936(46)$  \\
\hline
$48^3$
& LIN  & $-1.34(35) $ & $-1.12(22) $ & $-0.97(14) $ & $-0.84(11) $  \\
& EXP  & $-1.35(36) $ & $-1.13(22) $ & $-0.98(15) $ & $-0.85(11) $  \\
\end{tabular}
\end{center}
\caption{\label{SCL.table}
Our results for the scattering length $a_0/m_\pi$ ($1/{\rm GeV}^2$)
calculated from the energy shift
obtained by
the liner fitting (LIN) and
the exponential fitting (EXP) of $R_0(t)$ in $t$.
}
\end{table}
%
%
\begin{table}[t]
\begin{center}
\begin{tabular}{lrr}
                                    &  \multicolumn{1}{c}{Chiral}
                                    &  \multicolumn{1}{c}{No-Chiral}  \\
\hline
$A_{00}$                            & \multicolumn{1}{c}{---}
                                                    & $-0.069(41)$  \\
$A_{10}$  $( 1 / {\rm GeV}^2 )$     &  $-1.389(84)$ & $-1.01(24) $  \\
$A_{20}$  $( 1 / {\rm GeV}^4 )$     &  $ 0.79(18) $ & $ 0.33(33) $  \\
$A_{01}$  $( 1 / {\rm GeV}^2 )$     &  $-2.07(20) $ & $-2.00(20) $  \\
$A_{11}$  $( 1 / {\rm GeV}^4 )$     &  $ 3.22(47) $ & $ 3.09(48) $  \\
$A_{02}$  $( 1 / {\rm GeV}^4 )$     &  $ 1.27(53) $ & $ 1.23(53) $  \\
\hline
$\chi^2 / ND.$                      &  $ 0.863    $ & $0.782     $  \\
\end{tabular}
\end{center}
\caption{\label{A_fit.table}
Results of fitting of the scattering amplitude
with the assumption (Chiral), 
and without the assumption $A_{00}=0$ (No-Chiral).
}
\end{table}
%
%
\begin{table}[t]
\begin{center}
\begin{tabular}{ccr}
$p^2\ ({\rm GeV}^2)$ & $\sqrt{s}\ ({\rm GeV})$ & $\delta (p)\ ({\rm deg.})$   \\
\hline
$0.020$  &  $0.40$ &  $-2.71(12)$ \\
$0.070$  &  $0.60$ &  $-8.09(59)$ \\
$0.140$  &  $0.80$ &  $-14.8(12)$ \\
$0.230$  &  $1.00$ &  $-22.0(20)$ \\
$0.340$  &  $1.20$ &  $-28.6(31)$ \\
\end{tabular}
\end{center}
\caption{\label{D_LAT.table}
Our results for the scattering phase shift at several momenta at the physical pion mass.
}
\end{table}
%
%
%
\end{document}